\documentclass{webofc}
\usepackage[varg]{txfonts}   

\begin{document}
\title{Novel deep learning methods for track reconstruction}

\author{\firstname{Steven} \lastname{Farrell}\inst{1}\fnsep\thanks{\email{SFarrell@lbl.gov}} \and
        \firstname{Paolo} \lastname{Calafiura}\inst{1} \and
        \firstname{Mayur} \lastname{Mudigonda}\inst{1} \and
        \lastname{Prabhat}\inst{1} \and
        \firstname{Dustin} \lastname{Anderson}\inst{2} \and
        \firstname{Jean-Roch} \lastname{Vlimant}\inst{2} \and
        \firstname{Stephan} \lastname{Zheng}\inst{2} \and
        \firstname{Josh} \lastname{Bendavid}\inst{2} \and
        \firstname{Maria} \lastname{Spiropulu}\inst{2} \and
        \firstname{Giuseppe} \lastname{Cerati}\inst{3} \and
        \firstname{Lindsey} \lastname{Gray}\inst{3} \and
        \firstname{Jim} \lastname{Kowalkowski}\inst{3} \and
        \firstname{Panagiotis} \lastname{Spentzouris}\inst{3} \and
        \firstname{Aristeidis} \lastname{Tsaris}\inst{3}
}

\institute{Lawrence Berkeley National Laboratory \and
           California Institute of Technology \and
           Fermi National Accelerator Laboratory
}

\abstract{
For the past year, the HEP.TrkX project has been investigating machine learning solutions to LHC particle track reconstruction problems. A variety of models were studied that drew inspiration from computer vision applications and operated on an image-like representation of tracking detector data. While these approaches have shown some promise, image-based methods face challenges in scaling up to realistic HL-LHC data due to high dimensionality and sparsity. In contrast, models that can operate on the spacepoint representation of track measurements (“hits”) can exploit the structure of the data to solve tasks efficiently.
In this paper we will show two sets of new deep learning models for reconstructing tracks using space-point data arranged as sequences or connected graphs. In the first set of models, Recurrent Neural Networks (RNNs) are used to extrapolate, build, and evaluate track candidates akin to Kalman Filter algorithms. Such models can express their own uncertainty when trained with an appropriate likelihood loss function. The second set of models use Graph Neural Networks (GNNs) for the tasks of hit classification and segment classification. These models read a graph of connected hits and compute features on the nodes and edges. They adaptively learn which hit connections are important and which are spurious. The models are scaleable with simple architecture and relatively few parameters. Results for all models will be presented on ACTS generic detector simulated data.}

\maketitle

\section{Introduction}
\label{sec:intro}

Track reconstruction presents a challenging pattern recognition problem at the High Luminosity Large Hadron Collider (HL-LHC)~\cite{hllhc}. Collision events contain on average 200 interactions and O(10k) particles which leave O(100k) space-point ``hits'' in the detectors. Today's algorithms have trouble scaling to these conditions. The combinatorial seeding and track building algorithms are inherently serial and scale quadratic or worse with detector occupancy. It is thus worthwhile to investigate new solutions such as methods based on Deep Learning~\cite{deeplearning}.

The HEP.TrkX pilot project was founded to develop machine learning solutions for High Energy Physics (HEP) tracking.
Previous work was mainly on image-based methods such as image segmentation and image captioning~\cite{ctd2017}. While these methods were shown to have some success on simple toy data problems, they still face considerable challenges when scaling up to realistic HL-LHC conditions. In particular, realistic detector images have high dimensionality and sparsity, and are difficult to construct from the irregular geometries of HL-LHC detectors.

In contrast, the space-point representation of HEP tracking data presents some advantages along with new challenges. Models must be designed to consume data in the form of spatial arrangements of hits with variable sample size. But such models, when conceived, can potentially exploit the spatial structure of the data with the full precision of the detector.

In the rest of this paper we describe two classes of models that can operate on the space-point representation of tracking data: track building with Recurrent Neural Networks (RNNs) and track finding with Graph Neural Networks (GNNs).

The tracking data used in this paper is produced with the ACTS~\cite{acts} package using the barrel geometry of the generic HL-LHC detector. The simulated collision events consist of QCD physics interactions with a single hard-scatter process and a Poisson-distributed ($\mu=10$) number of additional soft QCD ``pileup'' interactions. Finally, in order to further simplify the problems, particles with transverse momentum below $1$ GeV are removed and duplicate hits on detector layers are removed.

\section{Track building with Recurrent Neural Networks}
\label{sec:rnnTracking}

Particle tracks can be represented as a sequence of hits which makes them applicable to Recurrent Neural Network architectures. In a similar way to how a Kalman Filter algorithm operates, we design RNN models which can model the dynamics of a particle trajectory and make extrapolated predictions.

The first model we present is an RNN which makes next-step hit predictions cast as a regression problem. The second model extends this by making predictions in the form of Gaussian probability distributions. Both models utilize Long Short Term Memory~\cite{lstm} networks for their powerful non-linear sequence modeling capabilities.

\subsection{Sequential hit predictor regression model}

\begin{figure}[htbp]
    \centering
    \includegraphics[width=0.6\textwidth]{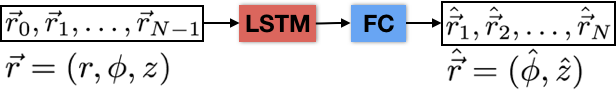}
    \caption{Diagram of the hit predictor model which takes a sequence of three-dimensional coordinates as input and produces two-dimensional next-step predictions. The architecture consists of an LSTM layer and a fully-connected layer.}
    \label{fig:rnnFilterModel}
\end{figure}

The RNN hit predictor model is illustrated in figure~\ref{fig:rnnFilterModel}. Given a sequence of hit coordinates, the model produces for every element a prediction of the position of the next hit conditioned on its position and the preceding hit positions. The learning problem is thus cast as a multi-target regression problem and the model is trained with a mean-squared-error loss function.

\begin{figure}[htbp]
    \centering
    \includegraphics[width=1\textwidth]{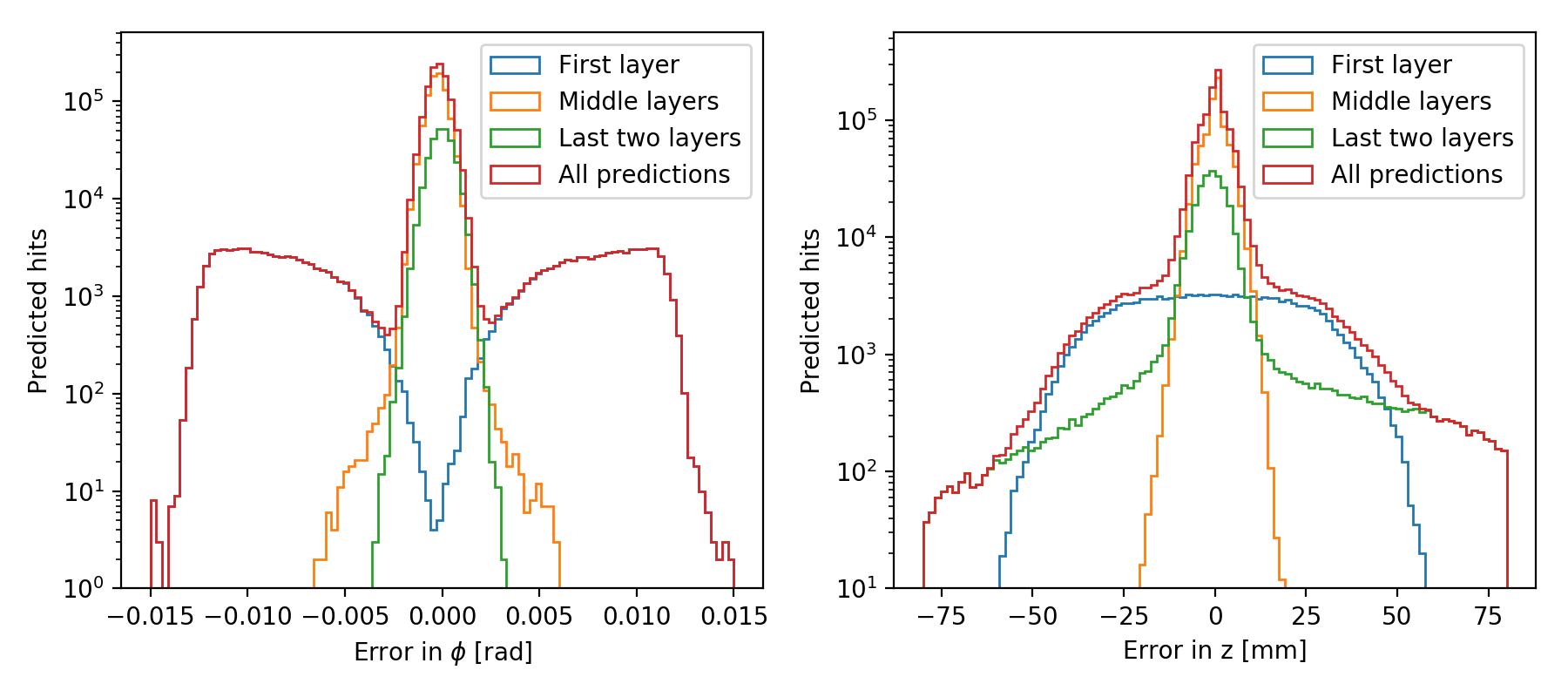}
    \caption{Residual errors for the hit predictor model in $\phi$ and $z$. The first layer has large errors but once the model sees two hits it is able to produce precise predictions for the remaining layers.}
    \label{fig:rnnFilterResidual}
\end{figure}

\begin{figure}[htbp]
    \centering
    \includegraphics[width=1\textwidth]{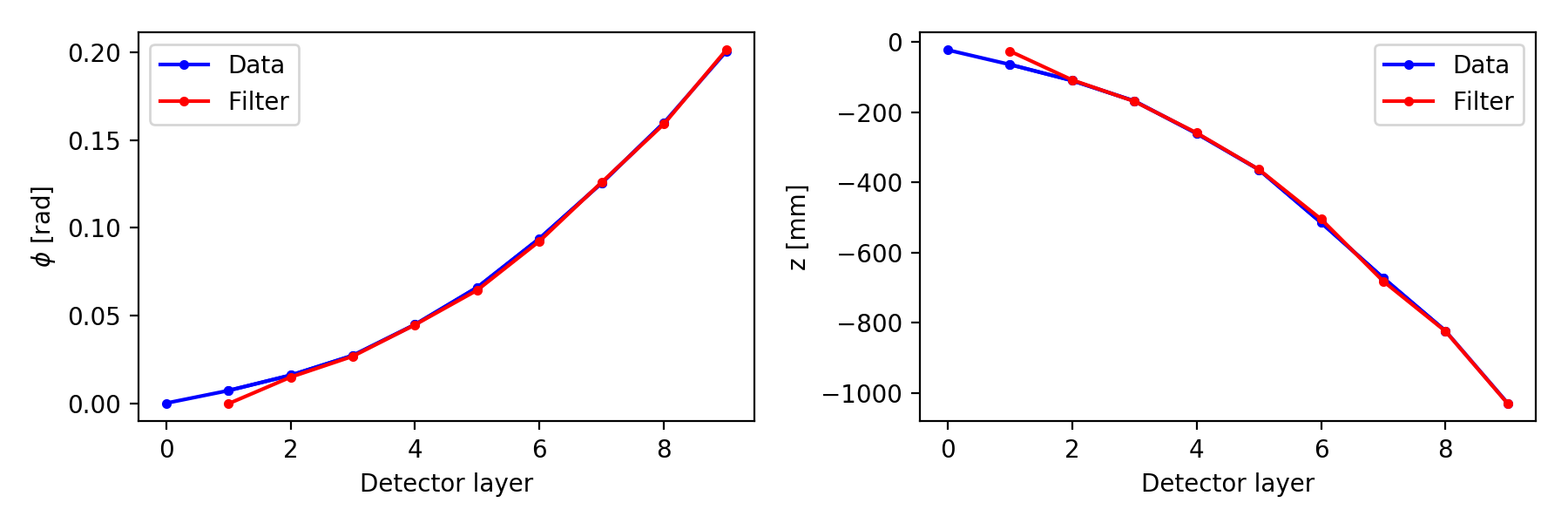}
    \caption{An example track with the hit predictor model's predictions.}
    \label{fig:rnnFilterTrajectory}
\end{figure}

The model is trained and evaluated on tracks that hit all ten barrel detector layers. The resulting residual errors evaluated on an independent test set are shown in figure~\ref{fig:rnnFilterResidual}. An example trajectory with the predictions is shown in figure~\ref{fig:rnnFilterTrajectory}. While the first prediction from the model is a poor guess because it cannot estimate the track trajectory, by the second hit in the sequence the model produces reasonable predictions. In the $z$ coordinate, most of the predictions fall within 1~mm.


\subsection{Sequential hit predictor Gaussian model}

The second RNN hit predictor model is nearly the same as the one described above, but it produces its predictions in the form of Gaussian probability distributions. The outputs of the network include the mean-value predictions as well as the parameters of a covariance matrix. The learning problem now is cast as a maximum likelihood estimation with the following Gaussian log-likelihood loss per hit prediction:
\begin{equation}
L(r, \hat{r}, \Sigma) = \log |\Sigma| + (r - \hat{r})^T \Sigma^{-1} (r - \hat{r})
\end{equation}
The advantage of this augmented approach is that the model learns to express its own uncertainty via the Gaussian distribution. This gives us a more interpretable, robust prediction which can be used to score candidate hits in track building.

\begin{figure}[htbp]
    \centering
    \includegraphics[width=1\textwidth]{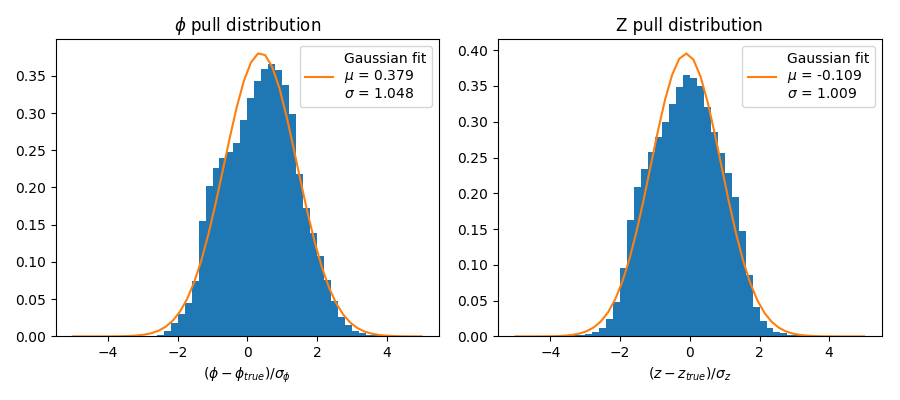}
    \caption{Pull distributions for the Gaussian hit predictor predictions with Gaussian fits. There are some clearly non-Gaussian effects in the pulls but the fitted width is consistent with one which means the model's uncertainty predictions are sensible.}
    \label{fig:rnnGausFilterResidual}
\end{figure}

\begin{figure}[htbp]
    \centering
    \includegraphics[width=1\textwidth]{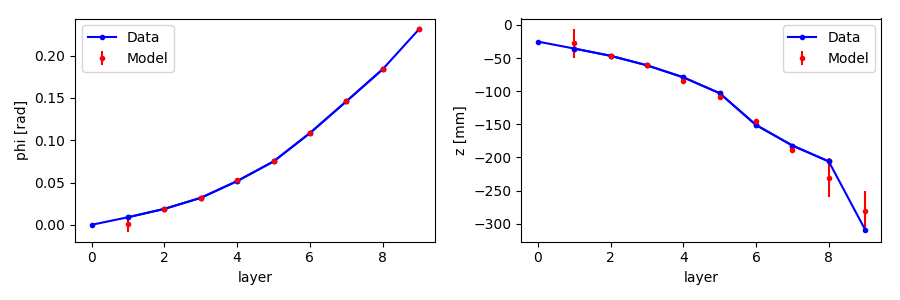}
    \caption{An example track and Gaussian hit predictor model's predictions with uncertainties. The predictions follow the track well and the uncertainties describe the trajectory wiggles in the coarse outer detector layers.}
    \label{fig:rnnGausFilterTrajectory}
\end{figure}

The resulting pull distributions of the predictions are shown in figure~\ref{fig:rnnGausFilterResidual} and an example trajectory with model predictions is shown in figure~\ref{fig:rnnGausFilterTrajectory}. This model also produces good predictions, though we do see some non-Gaussian features in the pull distributions which warrant further investigation in follow-up studies.


\begin{figure}[htbp]
    \centering
    \includegraphics[width=0.6\textwidth]{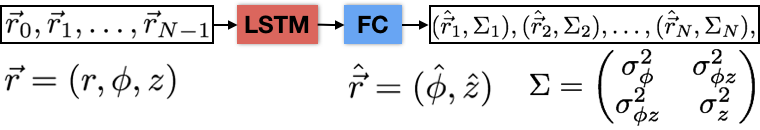}
    \caption{Diagram of the Gaussian hit predictor model which takes a sequence of three-dimensional coordinates as input and produces bi-variate Gaussian probability distributions as next-step predictions. The architecture is the same as the basic hit predictor but the model provides additional output which parameterizes the Gaussian covariance matrix.}
    \label{fig:rnnGauFilterModel}
\end{figure}

\subsection{Building tracks}

\begin{figure}[htbp]
    \centering
    \includegraphics[width=1\textwidth]{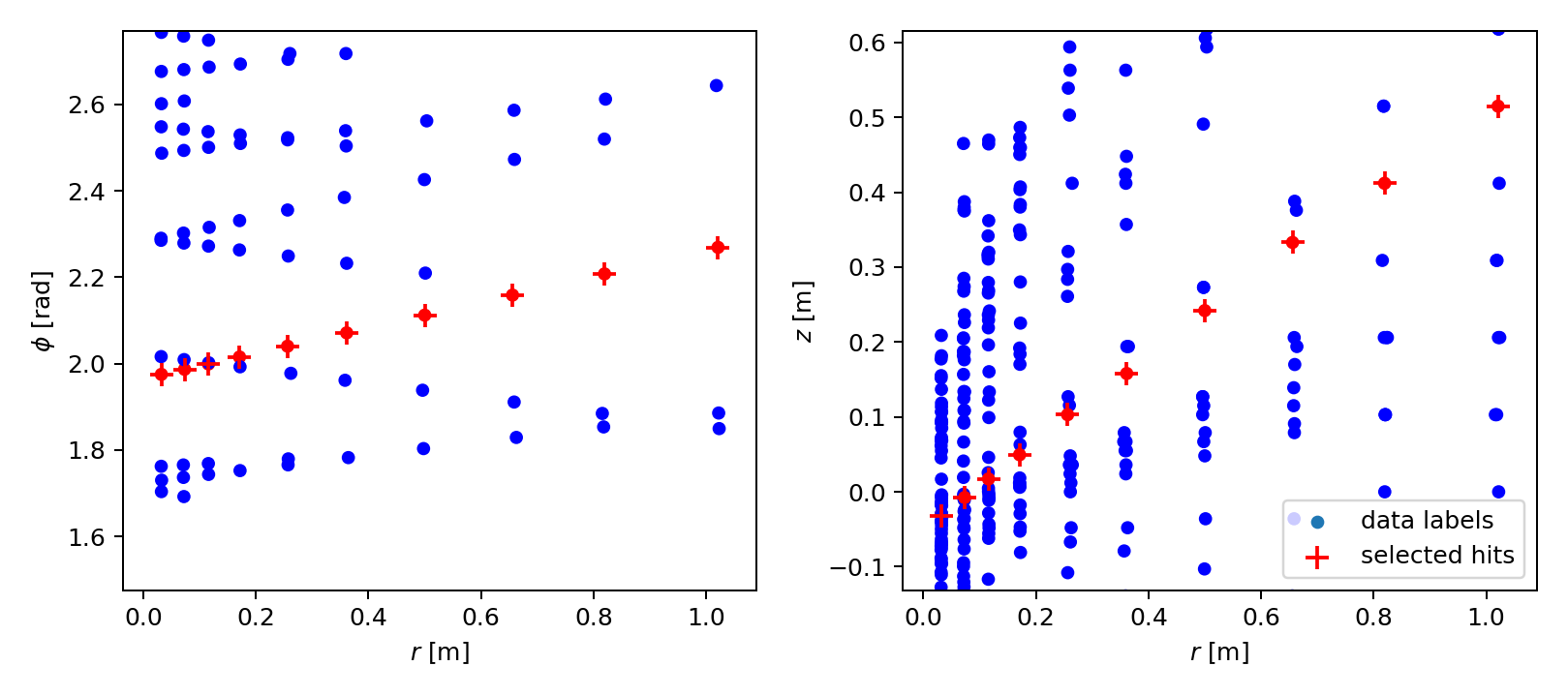}
    \caption{An example track properly reconstructed using the basic hit predictor model in an event.}
    \label{fig:rnnFilterTreeSearch}
\end{figure}

For a simple test of these models, we use them to extrapolate and build tracks in low-occupancy events. We construct a track ``seed'' using the initial three hits of a true track, then use the RNN models to make forward predictions and select the closest (or highest-scoring) hit in the event on each successive layer. An example track which is correctly fully reconstructed using the simple RNN hit predictor model is shown in figure~\ref{fig:rnnFilterTreeSearch}. In this simplified scenario both models are very good at making predictions for selecting candidate hits. The resulting hit selection accuracies measured are 99.93\% and 99.98\% for the simple and Gaussian models, respectively.

For a proper assessment of these models, a full combinatorial tree search algorithm with full occupancy collision data should be used. This is currently left for future work.


\section{Track finding with Graph Neural Networks}
\label{sec:gnnTracking}

\begin{figure}[htbp]
    \centering
    \includegraphics[width=0.5\textwidth]{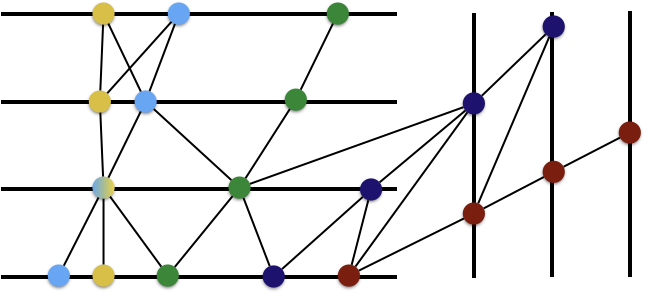}
    \caption{Illustration of a graph representation of track hit data. Hits are connected on adjacent layers if they are compatible according to some criteria.}
    \label{fig:hitGraph}
\end{figure}

Another way to represent tracking data with points is as a graph of connected hits. This is illustrated in figure~\ref{fig:hitGraph}. In this representation, we can apply a powerful class of methods from Geometric Deep Learning~\cite{gdl} known as Graph Neural Networks (GNNs). The graph can be constructed by connecting plausibly-related hits using geometric constraints or some kind of pre-processing algorithm like the Hough Transform. A GNN model can learn on this representation and solve tasks with predictions over the graph nodes, edges, or global state.

We have developed two applications using Graph Neural Networks. The first is a binary hit classification model which learns to identify one track in a partially-labeled graph by classifying the graph nodes. The second is a binary segment classification model which learns to identify many tracks at once by classifying the graph edges (hit pairs). The inputs to these models are the node features (the 3D hit coordinates) and the connectivity specification.

\subsection{Graph neural network architecture}

The architecture we have developed is similar to that of Interaction Networks~\cite{interaction-networks} but is customized for our purposes. Two main components operate locally on the graph:

\begin{itemize}
    \item An \textbf{EdgeNetwork} computes weights for every edge of the graph using the features of the start and end nodes.
    \item A \textbf{NodeNetwork} computes new features for every node using the edge weight aggregated features of the connected nodes on the previous and next detector layers separately as well as the nodes' current features.
\end{itemize}
Both the EdgeNetwork and NodeNetwork are implemented as Multi-Layer Perceptrons (MLPs) with two layers each and hyperbolic tangent hidden activations.

\begin{figure}[htbp]
    \centering
    \includegraphics[width=0.9\textwidth]{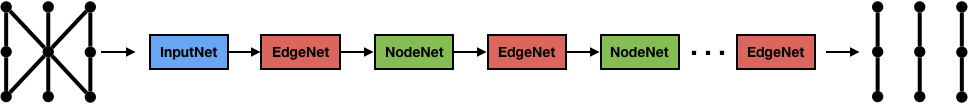}
    \caption{Diagram of the Graph Neural Network model which begins with an input transformation layer and has a number of recurrent iterations of alternating EdgeNetwork and NodeNetwork components. In this case, the final output layer is the EdgeNetwork, making this a segment classifier model.}
    \label{fig:gnnModel}
\end{figure}

The full Graph Neural Network model consists of an input transformation layer followed by recurrent alternating applications of the EdgeNetwork and NodeNetwork. The architecture for the segment classification network is illustrated in figure~\ref{fig:gnnModel}. With each iteration of the networks, the model propagates information through the graph, adaptively learning to strengthen important connections and weaken useless or spurious ones.

\subsection{Graph hit classification}

The hit classification model performs binary classification of the nodes of the graph using labels to specify three seed hits. The graphs are constructed by taking four hits on each detector layer in the region around the true track location and connecting all hits together on adjacent layers. The model uses seven graph iterations followed by a final classification layer with sigmoid activation that operates on every node to predict whether the nodes belong to the target track or not.

\begin{figure}[htbp]
    \centering
    \includegraphics[width=1\textwidth]{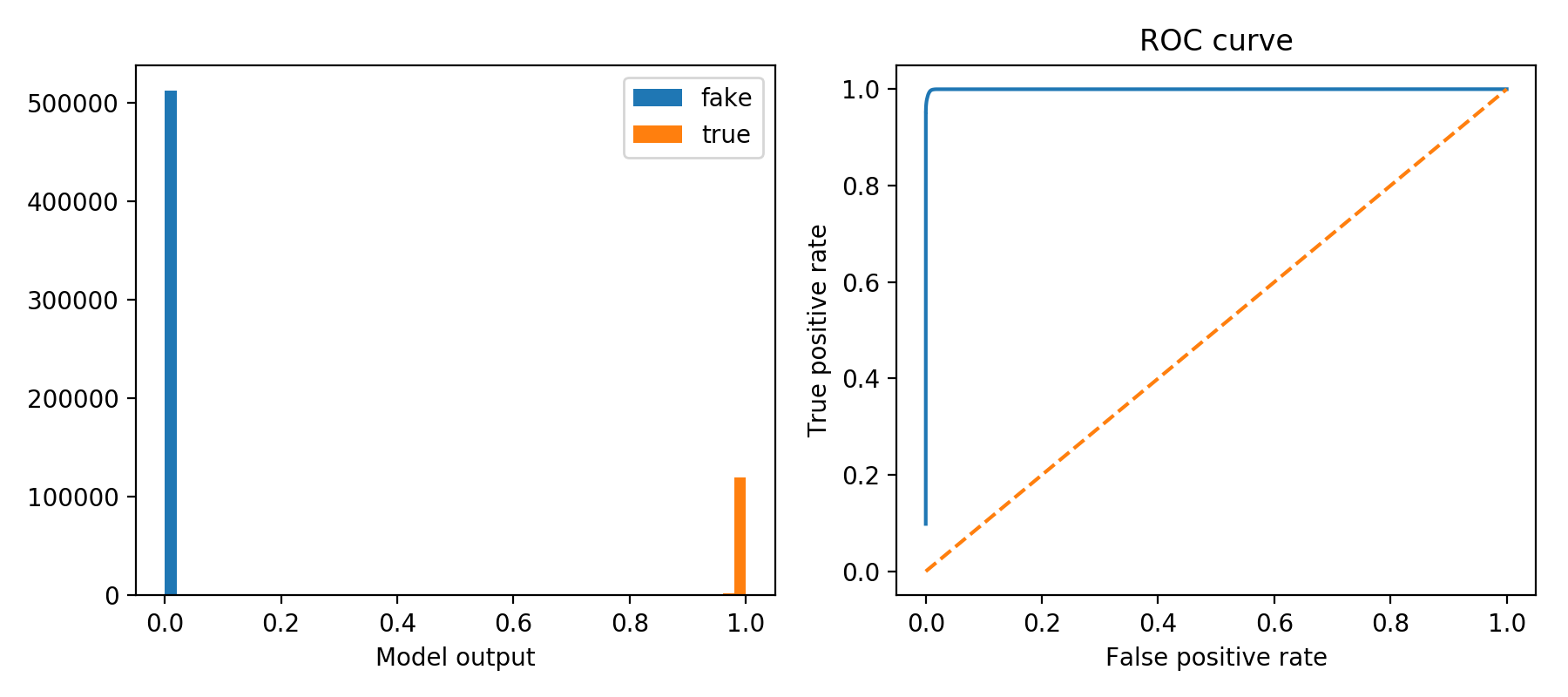}
    \caption{Output predictions and ROC curve for the GNN hit classification model. The outputs show perfect separation between real and fake hits, and the ROC curve shows excellent performance.}
    \label{fig:gnnHitPerformance}
\end{figure}

\begin{figure}[htbp]
    \centering
    \includegraphics[width=1\textwidth]{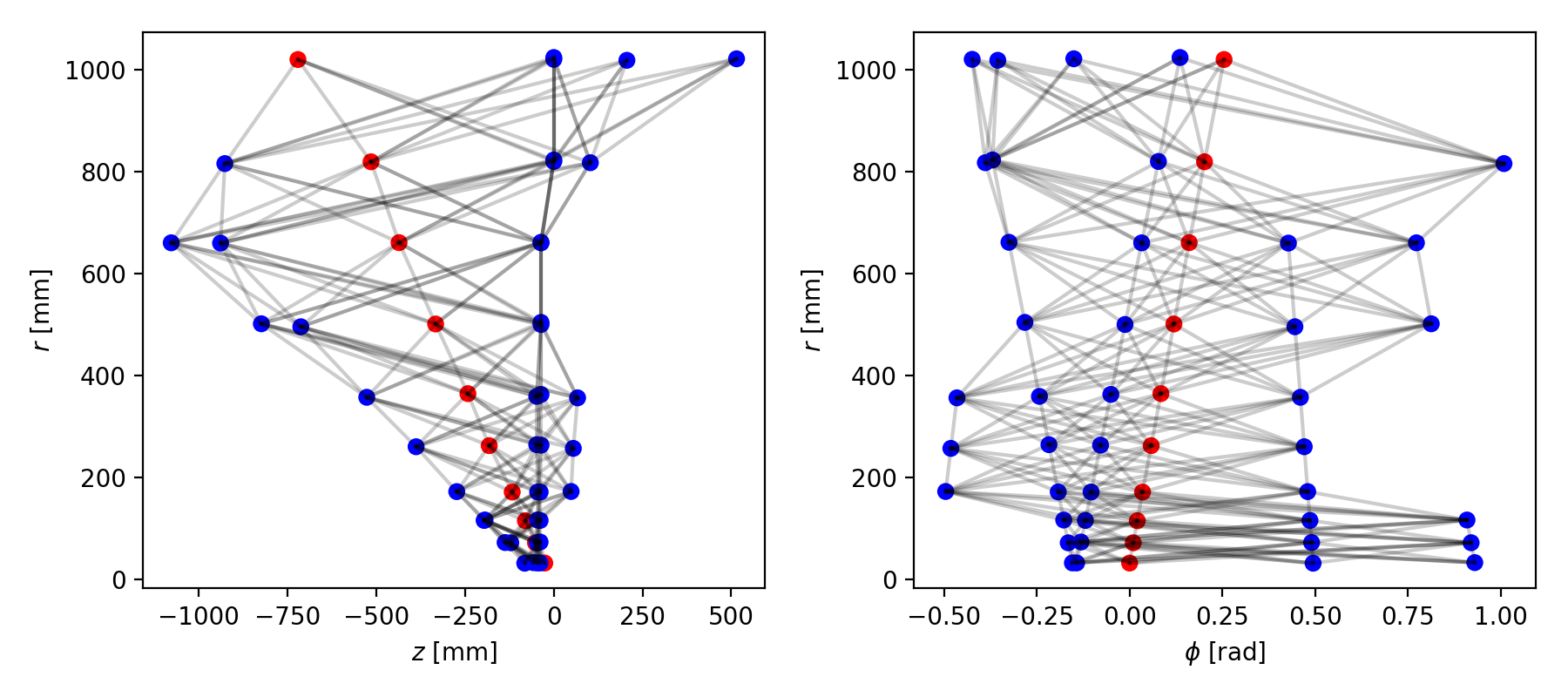}
    \caption{Example hit classification graph for a track. The colors indicate the model score, with red indicating 1 (correct hit) and blue indicating 0.}
    \label{fig:gnnHitExample}
\end{figure}

Results for the model are shown in figure~\ref{fig:gnnHitPerformance} and an example prediction is shown in figure~\ref{fig:gnnHitExample}. Using a threshold on the model score of 0.5 gives 99.2\% purity, 97.9\% efficiency, and overall accuracy of 99.4\%.

\subsection{Graph segment classification}

The segment classification model performs binary classification on the edges of the graph to distinguish true hit pairs (i.e., hits produced by the same particle) from spurious pairs. In this case, the graphs are constructed with geometric constraints. For this initial study, cuts on the difference in $\phi$ and $z$ coordinates of the hits of $\pi/4$ and $300$mm respectively were applied. The model uses four graph iterations and one final application of the EdgeNetwork to produce the output classification scores for every edge in the graph.

\begin{figure}[htbp]
    \centering
    \includegraphics[width=1\textwidth]{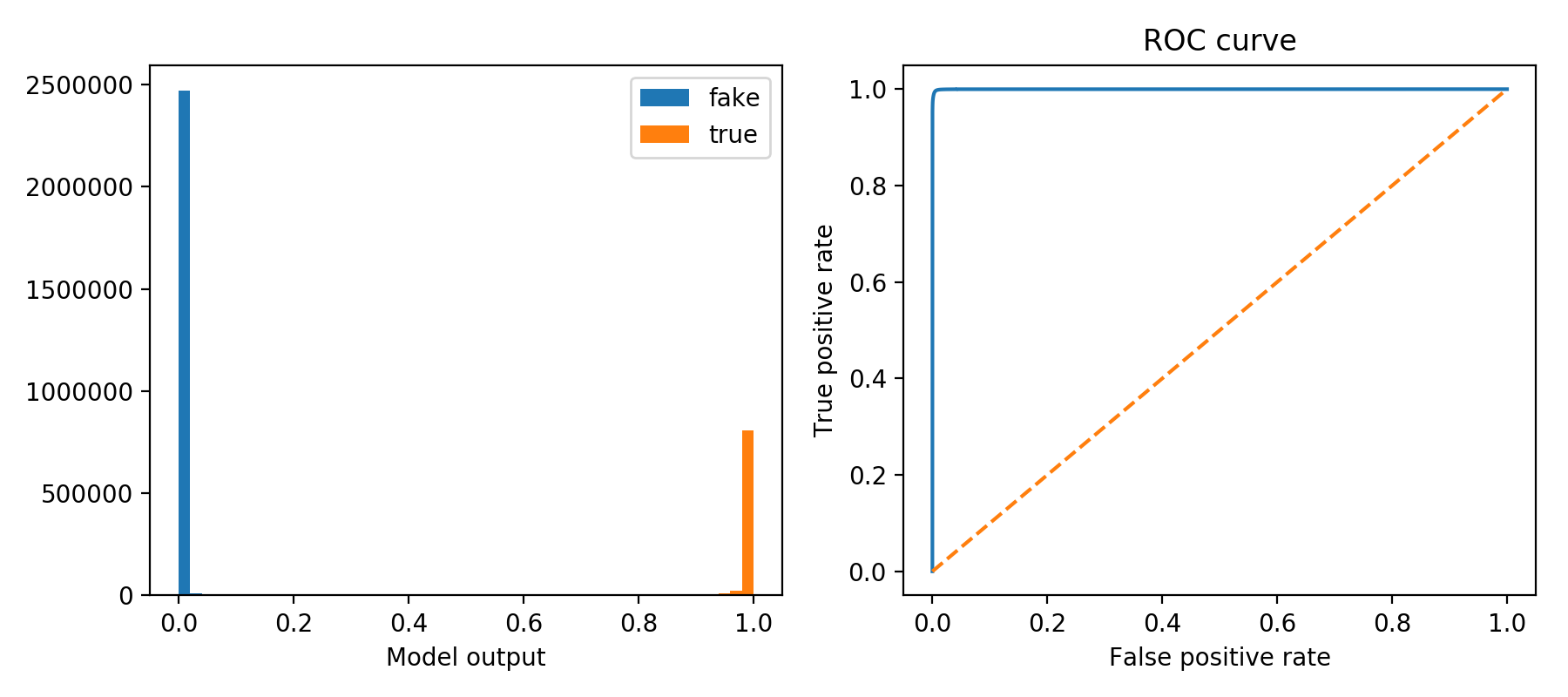}
    \caption{Output predictions and ROC curve for the GNN segment classification model. The outputs show perfect separation between real and fake segments, and the ROC curve shows excellent performance.}
    \label{fig:gnnSegPerformance}
\end{figure}

\begin{figure}[htbp]
    \centering
    \includegraphics[width=1\textwidth]{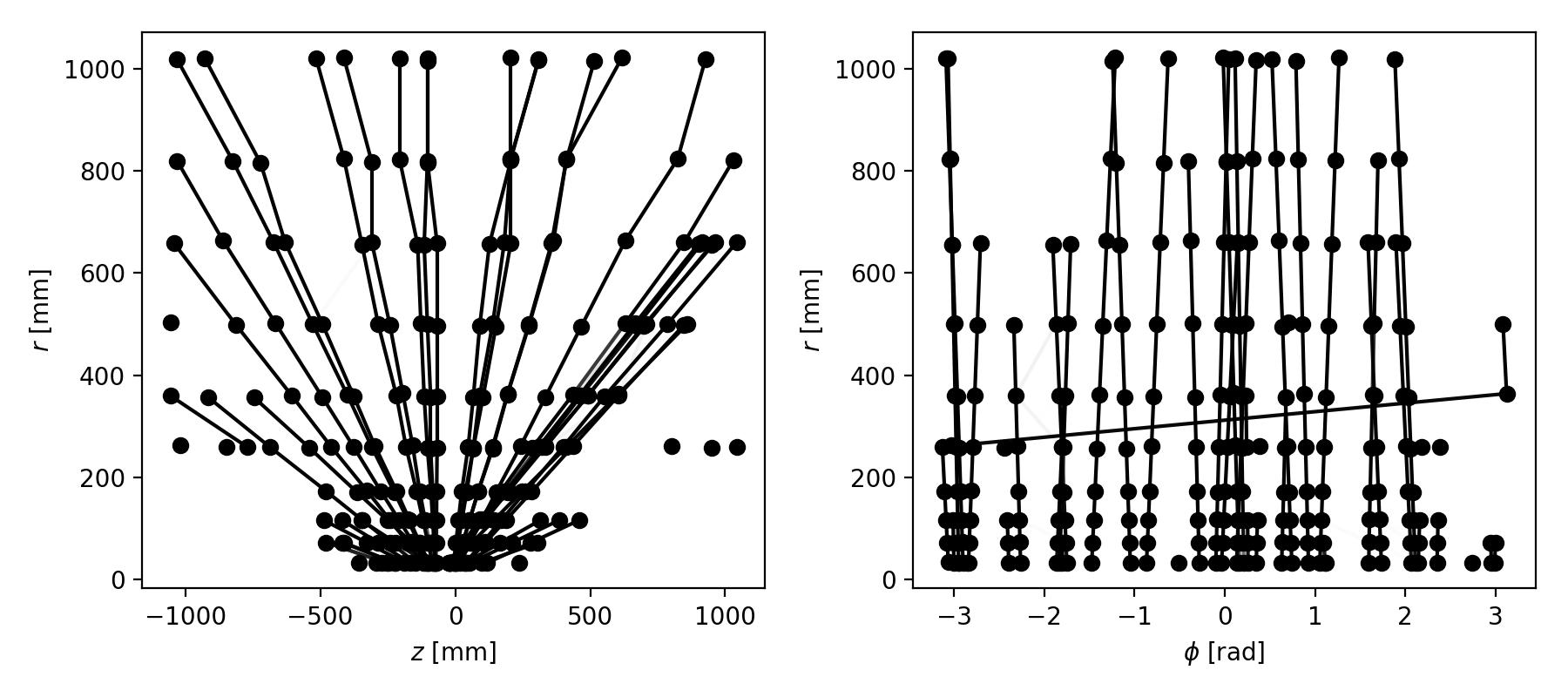}
    \caption{Example segment classification model output. Segments are drawn with opacity set according to the model output score. A segment with a value of 1 will be fully opaque and a segment with a value of 0 will be invisible. In this example, all segments specified in the input are classified correctly.}
    \label{fig:gnnSegExample}
\end{figure}

Results for the model are shown in figure~\ref{fig:gnnSegPerformance} and an example segment prediction is shown in figure~\ref{fig:gnnSegExample}. Using a threshold on the model score of 0.5 gives 99.5\% purity, 98.7\% efficiency, and overall accuracy of 99.5\%. These metrics are only defined relative to the segments constructed and presented to the model. The excellent performance of the model suggests we should be able to scale up complexity considerably.


\section{Conclusions}

We have demonstrated some novel methods for HEP track reconstruction based on deep learning. The first set of methods, based on Recurrent Neural Networks, demonstrate a capability similar to today's Kalman Filter algorithms for particle track state estimation and extrapolation. The second set of methods, based on Graph Neural Networks, show major promise for solving track reconstruction tasks by learning on an expressive structured graph representation of hit data. We believe the GNN approach to be our most promising deep learning solution for addressing the problems in tracking at the HL-LHC.

There is still work to be done to demonstrate the capability of these methods in a realistic tracking environment. The RNN track building methods need to be embedded into a complete combinatorial track tree search with robust candidate evaluation and compared with traditional combinatorial Kalman Filter algorithms on full occupancy detector data. The GNN methods need to be plugged into a complete tracking solution with robust, physics-driven graph construction methods as well as a method for the segment classifier which builds the final track candidates from the segment scores. The GNN methods also need to be scaled up to full data complexity in terms of occupancy, geometry (including detector endcaps), and features like missing or merged hits.

\section{Acknowledgements}

Part of this work was conducted at "\textit{iBanks}", the AI GPU cluster at Caltech. We acknowledge NVIDIA, SuperMicro  and the Kavli Foundation for their support of "\textit{iBanks}".


\bibliography{bibliography}

\begin{thebibliography}{7}

\bibitem{hllhc}
G.~Apollinari, O.~Brüning, T.~Nakamoto, L.~Rossi, CERN Yellow Report pp. 1--19
  (2015), \texttt{1705.08830}

\bibitem{deeplearning}
Y.~LeCun, Y.~Bengio, G.E. Hinton, Nature \textbf{521}, 436 (2015)

\bibitem{ctd2017}
{Farrell, Steven}, {Anderson, Dustin}, {Calafiura, Paolo}, {Cerati, Giuseppe},
  {Gray, Lindsey}, {Kowalkowski, Jim}, {Mudigonda, Mayur}, {Prabhat},
  {Spentzouris, Panagiotis}, {Spiropoulou, Maria} et~al., EPJ Web Conf.
  \textbf{150}, 00003 (2017)

\bibitem{acts}
\emph{{A Common Tracking Software Project}},
  \urlstyle{tt}\url{http://acts.web.cern.ch/ACTS/index.php}

\bibitem{lstm}
S.~Hochreiter, J.~Schmidhuber, Neural Comput. \textbf{9}, 1735 (1997)

\bibitem{gdl}
M.M. Bronstein, J.~Bruna, Y.~LeCun, A.~Szlam, P.~Vandergheynst, CoRR
  \textbf{abs/1611.08097} (2016), \texttt{1611.08097}

\bibitem{interaction-networks}
P.W. Battaglia, R.~Pascanu, M.~Lai, D.J. Rezende, K.~Kavukcuoglu, CoRR
  \textbf{abs/1612.00222} (2016), \texttt{1612.00222}

\end{thebibliography}
%
%
%
%

\end{document}